\documentclass[12pt]{article}

\pdfoutput=1

\usepackage{latexsym,amsfonts,bm,epsfig,amsmath,natbib,authblk,amsthm,thmtools,amssymb}

\usepackage{float}
\usepackage{booktabs}
\usepackage{graphicx}
\usepackage[margin=.5cm]{caption}
\usepackage{xr-hyper}
\usepackage{color}
\usepackage[colorlinks,linkcolor=black,citecolor=black,filecolor=black,bookmarks=false,pagebackref]{hyperref}

\usepackage{pgfplots}
\usepackage{tikz}
\newlength\figureheight
\newlength\figurewidth

\floatstyle{plain}
\newfloat{Algorithm}{thp}{lop}
\floatname{Algorithm}{Algorithm}

\usepackage[plain,noend]{algorithm2e}

\makeatletter
\renewcommand{\algocf@captiontext}[2]{#1\algocf@typo. \AlCapFnt{}#2} 
\def\@algocf@capt@plain{top}
\renewcommand{\algocf@makecaption}[2]{%
  \addtolength{\hsize}{\algomargin}%
  \sbox\@tempboxa{\algocf@captiontext{#1}{#2}}%
  \ifdim\wd\@tempboxa >\hsize
   \hskip .5\algomargin%
    \parbox[t]{\hsize}{\algocf@captiontext{#1}{#2}}
 \else%
   \global\@minipagefalse%
   \hbox to\hsize{\box\@tempboxa}
 \fi%
  \addtolength{\hsize}{-\algomargin}%
}
\makeatother

\newcommand{\ignore}[1]{}

\newcommand{\y}{{y}}
\newcommand{\z}{z}
\newcommand{\dt}{\mathrm{d}}
 
\newcommand{\E}{\mathbb{E}}

\theoremstyle{remark}

\begin{document}
\def\spacingset#1{\renewcommand{\baselinestretch}%
	{#1}\small\normalsize} \spacingset{1}

\title{Bayesian Synthetic Likelihood}
\date{\empty}
\author[1]{David T. Frazier\thanks{Corresponding author:  david.frazier@monash.edu}}
\author[2]{Christopher Drovandi}
\author[3,4]{David J. Nott}
\affil[1]{Department of Econometrics and Business Statistics, Monash University, Clayton VIC 3800, Australia}
\affil[3]{Department of Statistics and Applied Probability, National University of Singapore, Singapore 117546}
\affil[4]{Operations Research and Analytics Cluster, National University of Singapore, Singapore 119077}
\affil[2]{School of Mathematical Sciences, Queensland University of Technology, Brisbane 4000 Australia}

\maketitle

\begin{abstract}
Bayesian statistics is concerned with conducting posterior inference for the unknown quantities in a given statistical model.  Conventional Bayesian inference requires the specification of a probabilistic model for the observed data, and the construction of the resulting likelihood function. However, sometimes the model is so complicated that evaluation of the likelihood is infeasible, which renders exact Bayesian inference impossible. 
	Bayesian synthetic likelihood (BSL) is a posterior approximation procedure that can be used to conduct inference in situations where the likelihood is intractable, but where simulation from the model is straightforward. In this entry, we give a high-level presentation of BSL, and its extensions aimed at delivering scalable and robust posterior inferences. 
\vspace{1cm}

\end{abstract}
\spacingset{1.9} 

\section{Introduction \label{sec:introduction}}
Consider that we wish to model observed data $y=(y_1,\dots,y_n)$ using a parametric model which we describe using its distribution function $P^{(n)}_\theta$, and which depends on unknown parameters
$\theta\in \Theta\subseteq\mathbb{R}^{d_\theta}$.  
The density of $P^{(n)}_\theta$ will be written $p^{(n)}_\theta(y)$, and for observed data $y$, the function $\theta\mapsto p^{(n)}_\theta(y)$ denotes the likelihood.  Given a prior density over the unknown parameters, denoted by $\pi(\theta)$, we can conduct 
Bayesian inference using the posterior density  
$$
\pi(\theta\mid y)= \frac{p^{(n)}_\theta(y)\pi(\theta)}{\int_\Theta p^{(n)}_\theta(y)\pi(\theta)\dt\theta}.
$$
In Bayesian applications, the posterior is often summarized using Monte Carlo
methods, such as Markov chain Monte Carlo (MCMC) (see \citealp{brooks2011handbook} for a handbook reference, and \citealp{roberts2017mcmc} for an encyclopedia entry).  MCMC can be applied in a wide variety of cases to produce samples from the exact posterior  $\pi(\theta\mid y)$, but these algorithms require that we can evaluate the likelihood or estimate
it unbiasedly.

When the likelihood has no closed-form expression, or is otherwise intractable, ``likelihood-free" methods are a common approach for conducting Bayesian inference.  While approximate Bayesian computation (ABC) \citep{sisson2018handbook,drovandi2017approximate} is perhaps the most well-established likelihood-free method, there 
are many other approaches.  Most likelihood-free algorithms share the idea
of using model simulation to circumvent the intractability of the likelihood:  draws from an approximate posterior can be found using algorithms that select samples for $\theta$ which generate synthetic data $\z=(z_1,\dots, z_n)^\top\sim P_\theta^{(n)}$ that is ``close enough'' to the observed data $\y$. 

In high-dimensional cases, directly matching noisy synthetic and observed data
can result in unreliable inferences.  Careful construction of a distance measure
can help. Consequently, the data is often reduced to a low-dimensional vector of summary statistics via a mapping $S:\mathbb{R}^{n}\rightarrow\mathcal{S}\subseteq\mathbb{R}^{d_s}$, where $d_s\ge d_\theta$, and then the distance between the observed and
synthetic data is constructed in the summary statistic space.   If variation irrelevant to $\theta$ is removed through the summary statistic mapping, 
this can improve the performance of common likelihood-free algorithms.  In what follows we show dependence of the summary on $n$ explicitly, and write $S_n(\cdot)$ for the summary mapping,  and $S_n=S_n(y)$ for the summary for the observed data $y$ when no confusion will result.  

While reduction of $y$ to the summary $S_n(y)$ can improve the performance of
likelihood-free algorithms, there is a loss of information unless $S_n(y)$ is sufficient.  Moreover, in most cases the models to which likelihood-free methods are applied are outside the exponential family, so that a set of sufficient statistics with fixed-dimension will not exist (see, \citealp{arnold2014sufficient} for a handbook entry on sufficient statistics). Nonetheless, since the likelihood is intractable, we may be willing to sacrifice some information in order to feasibly perform inference on the unknowns. 

If the likelihood of the summaries, $p(S_n\mid\theta)$, were tractable, then inference on $\theta$ could be conducted using the exact partial posterior
$$
\pi(\theta\mid S_n)=\frac{p(S_n\mid\theta)\pi(\theta)}{\int_\Theta p(S_n\mid\theta)\pi(\theta)\dt\theta},
$$which is referred to as the partial posterior since it only conditions on part of the data, $S_n$, and not the entire data series $y$. Unfortunately,  $p(S_n\mid\theta)$ is unlikely to be tractable, since the likelihood $p^{(n)}_\theta(y)$ is intractable.


The traditional ABC approach to likelihood-free inference  replaces the summary likelihood $p(S_n\mid\theta)$ with the simulation-based estimate
\begin{equation}\label{eq:peps}
p_{\epsilon}(S_n\mid\theta)=\frac{1}{M}\sum_{i=1}^{M}K_\epsilon(\|S_n(y)-S_n(z^i)\|),
\end{equation}
where $K_\epsilon(\cdot)$ is a known kernel function with bandwidth $\epsilon$, $z^i|\theta \stackrel{iid}{\sim} P_\theta^{(n)}$, and $M\ge1$ is chosen by the user. The use of the likelihood estimator $p_{\epsilon}(S_n\mid\theta)$ within sampling for $\theta$ produces draws from the ABC posterior
$$
\pi_{\mathrm{A}}(\theta\mid S_n)=\frac{\E_{z^1,\dots,z^M}p_{\epsilon}(S_n\mid\theta)\pi(\theta) }{\int_\Theta \E_{z^1,\dots,z^M}p_{\epsilon}(S_n\mid\theta\pi(\theta)\dt\theta}=\frac{\int K_\epsilon(\|S_n(y)-S_n(z)\|) \dt P^{(n)}_\theta(z)\pi(\theta)}{\int_\Theta \int K_\epsilon(\|S_n(y)-S_n(z)\|) \dt P^{(n)}_\theta(z)\pi(\theta)\dt\theta},
$$where the second equality follows since $p_{\epsilon}(S_n\mid\theta)$ is an unbiased estimator of $$\int K_\epsilon(\|S_n(y)-S_n(z)\|) \dt P^{(n)}_\theta(z),$$ for any $M\ge1$. 

Consequently, the ABC posterior $\pi_{\mathrm{A}}(\theta\mid S_n)$ attempts to estimate the likelihood of the summaries $p(S_n\mid\theta)$ with the conditional density estimator $p_{\epsilon}(S_n\mid\theta)$. As with all conditional density estimation (see \citealp{scott2014kernel} for a handbook entry), the ABC likelihood estimator suffers from the curse-of-dimensionality.  Even if the dimension of $S_n$ is moderate, the resulting ABC posterior $\pi_{\mathrm{A}}(\theta\mid S_n)$ can be a poor approximation to the partial posterior $\pi(\theta\mid S_n)$; we refer to \cite{blum2010non}, as well as \cite{FMRR2016}, and \cite{LF2016} for detailed discussion of this point.

\section{Bayesian Synthetic Likelihood}

Motivated by the fact that well-chosen summaries $S_n$ will satisfy a central limit theorem, rather than estimate the likelihood as in ABC, \cite{wood2010statistical} proposed to approximate the summary likelihood $p(S_n \mid \theta)$ using a Gaussian distribution with mean $b_{}(\theta)=\mathbb{E}\{S_n(z)\mid \theta\}$ and variance $\Sigma_n(\theta)=\text{var}\{S_n(z)\mid \theta\}$. \cite{wood2010statistical} refers to $N\{S_n;b(\theta),\Sigma_n(\theta)\}$ as the synthetic likelihood, where $N(x;\mu,\Sigma)$ denotes a Gaussian density function evaluated at $x$ with mean $\mu$ and variance matrix $\Sigma$. 

Since $b(\theta)$ and $\Sigma_n(\theta)$ are unlikely to be known in practice for any $\theta$, \cite{wood2010statistical} estimates these quantities using sample versions calculated from $M$ independent and identically distributed samples $z^i$, $i=1,\dots, M$ from $P_\theta^{(n)}$; $b(\theta)$ is estimated using the sample mean $\overline{b}_M(\theta)=M^{-1}\sum_{i=1}^{M}S_n(z^i)/M$, and $\Sigma_n(\theta)$ is estimated using the sample covariance $\overline{\Sigma}_M(\theta)=(M-1)^{-1}\sum_{i=1}^{M}\{S_n(z^i)-\overline{b}_M(\theta)\}\{S_n(z^i)-\overline{b}_M(\theta)\}$.  The estimates replace the unknown $b(\theta)$ and $\Sigma_n(\theta)$ in the synthetic likelihood, resulting in a noisy estimate $N\{S_n;\overline{b}_M(\theta),\overline{\Sigma}_M(\theta)\}$. 

While \cite{wood2010statistical} proposed synthetic likelihood, \cite{price2018bayesian} were the first to consider in detail its use for Bayesian inference.  \cite{price2018bayesian} study the repercussions of replacing the exact synthetic likelihood $N\{S_n;b(\theta),\Sigma_n(\theta)\}$, by its noisy counterpart $N\{S_n;\overline{b}_M(\theta),\overline{\Sigma}_M(\theta)\}$ inside Metropolis-Hastings MCMC (MH-MCMC) algorithms (see \citealp{RobertMHMCMC} for an encyclopedia entry on the MH-MCMC algorithm). \cite{price2018bayesian} demonstrate that using the noisy synthetic likelihood $N\{S_n;\overline{b}_M(\theta),\overline{\Sigma}_M(\theta)\}$ within MCMC results in a pseudo-marginal MCMC scheme (\citealp{andrieu2009pseudo}) that delivers draws from the following target posterior density
\begin{flalign*}
	\overline{\pi}_{\mathrm{B}}(\theta\mid S_n)&\propto {\pi(\theta)\overline{g}_n(S_n\mid\theta)}
	,
	\\	\overline{g}_n(S_n\mid\theta) &= \int N\{S_n;\overline{b}_M(\theta),\overline{\Sigma}_M(\theta)\} \prod_{i=1}^M \dt P^{(n)}_\theta\{S_n(\z^i)\}\,\dt S_n(\z^1)\,\dots\, \dt S_n(\z^M),
\end{flalign*}where $\overline{g}_n(S_n\mid\theta)$ is the expectation of $N\{S_n;\overline{b}_M(\theta),\overline{\Sigma}_M(\theta)\}$ with respect to the synthetic
data $z^1,\dots, z^M$.

 In contrast to the BSL posterior $\overline{\pi}_{\mathrm{B}}(\theta\mid S_n)$, if it were feasible to use the exact synthetic likelihood, we could obtain samples from the posterior density target
\begin{equation*}
	{\pi}_{\mathrm{B}}(\theta\mid S_n)\propto{\pi(\theta)N\{S_n;{b}(\theta),{\Sigma}_n(\theta)\}},
\end{equation*}which is referred to as the exact BSL posterior, while  $\overline{\pi}_{\mathrm{B}}(\theta\mid S_n)$ is called the BSL posterior. While $\overline{\pi}_{\mathrm{B}}(\theta\mid S_n)$ and ${\pi}_{\mathrm{B}}(\theta\mid S_n)$ differ in general, \cite{frazier2019bayesian} establish that the BSL posterior $\overline{\pi}_{\mathrm{B}}(\theta\mid S_n)$ can be arbitrarily close to the exact BSL posterior ${\pi}_{\mathrm{B}}(\theta\mid S_n)$ if we choose $M$ large enough.

While the BSL posterior $\overline{\pi}_{\mathrm{B}}(\theta\mid S_n)$ cannot be analytically calculated, it is possible to obtain samples from $\overline{\pi}_{\mathrm{B}}(\theta\mid S_n)$ using existing, and efficient, MCMC algorithms; with MH-MCMC, or its adaptive variants, being the most commonly applied approach.

Since its inception, BSL has been used in many different applications.   For example, it has been applied to a stochastic model of spreading of melanoma cells in \citet{Drovandi2018}.  \citet{Hartig2014} apply synthetic likelihood to the FORMIND model, an individual-based model of tropical forest dynamics.  \citet{Brown2014} employ synthetic likelihood for calibrating a stochastic model of Avian flu transmission, and \citet{Biswas2022} estimate the rate parameters of a stochastic kinetic model in genetics using BSL.

\subsection{Comparison between ABC and BSL}\label{sec:agree} 
ABC and BSL are both likelihood-free methods that seek to conduct posterior inference in situations where the likelihood is intractable. In the case of correctly specified models, \cite{frazier2019bayesian} demonstrate that ABC and BSL have similar theoretical behavior, with both methods displaying certain optimality properties that we desire in Bayesian methods. 

In addition, the authors show that if the asymptotic normality assumption for $S_n(y)$ and $S_n(z)$ is reasonable, then the BSL posterior can have higher sampling efficiency than the ABC posterior. Consequently, BSL can partly alleviate the curse of dimensionality that is often associated with ABC-based inference (\citealp{blum2010non}, \citealp{FMRR2016}).

In contrast to ABC, which must choose the tolerance $\epsilon$ used to estimate the likelihood of the summaries, BSL does not require such a tuning parameter. In BSL, all we must choose is the number of simulated datasets, $M,$ employed to estimate the unknown mean and variance of the summaries ($b(\theta)$ and $\Sigma_n(\theta)$). In general, the choice of $M$ does not have a large impact on the resulting behavior of the BSL posterior,\footnote{\cite{price2018bayesian} demonstrate empirically that the choice of $M$ used to estimate the mean and variance often has little impact on the behavior of the posterior (this is verified theoretically in the analysis of \citealp{frazier2019bayesian}).} although it may impact the performance of sampling algorithms.  Moreover, since the BSL posterior is based on a Gaussian approximation, rather than a nonparametric estimator of the likelihood as in ABC, the BSL posterior scales more readily to higher-dimensional parameters and summaries than ABC methods.   \citet{frazier2019bayesian} formalize this superior scaling behaviour for basic
ABC algorithms, and demonstrate that BSL performs similarly to ABC
approaches with linear regression adjustment \citep{Beaumont2025}.

\subsection{Extensions of BSL: Flexibility and Sampling Efficiency}

There have been a number of extensions to the standard BSL procedure just described, with the motivation to improve its flexibility or to increase its sampling efficiency (we discuss the extension to model misspecification in Section 2.3).  In terms of flexibility, \citet{Fasiolo2018} use a saddlepoint approximation while \citet{an2020robust} use kernel density estimation (KDE) to model the summary marginal distributions and a Gaussian copula to model the dependence between summaries.  \citet{Priddle2020} extend \citet{an2020robust} to implement transformation KDE \citep{Wand1991} within BSL, which adapts the hyperbolic power transformation of \cite{Tsai2017} to account for heavier skewness and/or kurtosis in the distributions of marginal summaries.

In terms of computational efficiency, \citet{priddle2019efficient} show that to control the variance of the synthetic likelihood estimator, and to help ensure that MCMC mixes reasonably well, the number of model simulations $M$ must scale at least as $M = \mathcal{O}(d_s^2)$ with the dimension of the summary.  \citet{priddle2019efficient} exploit a whitening transformation to decorrelate the summaries producing a diagonal covariance matrix, which then only requires $M = \mathcal{O}(d_s)$ model simulations to control the variance.  Since the whitening transformation only decorrelates the model summaries at some central parameter value, \citet{priddle2019efficient} consider using the covariance shrinkage estimator of \citet{Warton2008}:
\begin{align*}
	\bar{\Sigma}_{\gamma} = \bar{\Sigma}_{d,M}^{1/2}(\gamma \bar{R}_M + (1-\gamma)I_{d_s})\bar{\Sigma}_{d,M}^{1/2},
\end{align*} 
where $\bar{R}_M$ is the correlation matrix and $\bar{\Sigma}_{d,M}$ is the diagonal matrix of variances, all estimated using $M$ model simulations for a proposed value of $\theta$ within MCMC.  Thus, $\bar{\Sigma}_{\gamma}$ shrinks towards a diagonal covariance matrix, but not all the way to accommodate any correlation between summaries not accounted for by the whitening transformation.  For a related approach that exploits the graphical lasso, see \citet{An2019}. {Many of the extensions listed above, together with the standard BSL approach, are implemented in the \texttt{BSL} \texttt{R} package \citep{an2019bsl} and within \texttt{ELFI} (engine for likelihood-free inference, \citealp{Lintusaari2018}) in \texttt{Python}.} 

Another way to speed up BSL is to resort to a variational approximation (see \citealp{tranVB} for a handbook entry on variational methods) as opposed to MCMC, replacing sampling with optimisation.  The first variational BSL (VBSL) approach is developed in \citet{Ong2018}, which demonstrates that significantly fewer model simulations are required compared to MCMC, at the expense of a Gaussian approximation to the BSL posterior.  \citet{Ong2018a} then extend VBSL so that it can be applied to high-dimensional problems, by exploiting the shrinkage estimator above, $\bar{\Sigma}_{\gamma}$, control variates and a factor structure of the posterior distribution.

There are also other approaches for accelerating BSL.  One approach is to use a surrogate model, such as a Gaussian process (GP).  \citet{Meeds2014} use a GP surrogate for each summary and approximate the likelihood assuming that the summaries are independent.  \citet{Wilkinson2014} use a GP model of the synthetic likelihood itself.  Once the GP surrogates are sufficiently well trained, no additional model simulations are required.  As an alternative approach, \citet{Levi2022} develop a method for recycling model simulations from the MCMC chains' history within ABC and BSL. \citet{Picchini2022} develop a proposal for MCMC that conditions on the observed data to speed-up convergence, as well as exploiting ideas from the correlated pseudo-marginal literature (e.g.\ \citealp{Tran2016}) to reduce the number of model simulations required to estimated the synthetic likelihood. {An alternative to both ABC, and BSL is sequential neural likelihood (\citealp{Papamakarios2019}) methods, which aims to be simultaneously more flexible and use the model simulations in a more efficient manner than BSL.  Given space restrictions, we do not discuss this, or related alternative methods. }

\subsection{Extensions of BSL: Model Misspecification}
BSL seeks to conduct posterior inference in models that are so complex that the resulting likelihood function is not analytically available. In such cases, it may be unwise to assume that the model is a faithful representation of the data-generating process. In the context of likelihood-free inference methods, like ABC and BSL, model misspecification is taken to mean that there is no value of $\theta\in\Theta$ such that the observed summaries $S_n(y)$ can be matched by the simulated summaries $S_n(z)$ (see \citealp{frazier2020model} for a formal definition). Critically, it has been shown that the agreement between ABC and BSL methods discussed in Section \ref{sec:agree} breaks down when the model used to simulate synthetic data $z$, i.e., $P_\theta^{(n)}$, is misspecified. In such cases, \cite{frazier2021synthetic} demonstrate that the two methods can deliver surprisingly different posterior inferences, and that both methods can be poorly behaved. 

We can trace the differences in ABC and BSL posteriors in misspecified models to the choice of `likelihood' that is used in their construction. In the case of ABC, the likelihood used is $p_\epsilon(S_n\mid\theta)$ in \eqref{eq:peps}, which depends explicitly on the kernel-distance $K_\epsilon(\|S_n(y)-S_n(z)\|)$, and for BSL, we use the noisy synthetic likelihood $N\{S_n;\overline{b}_M(\theta),\overline{\Sigma}_M(\theta)\}$. Clearly, there is no reason to believe \textit{a priori} that values of $\theta$ that make $p_\epsilon(S_n\mid\theta)$ large are related to values of $\theta$ that make $N\{S_n;\overline{b}_M(\theta),\overline{\Sigma}_M(\theta)\}$ large; with such values determining where $\pi_{\mathrm{A}}(\theta\mid S_n)$ and $\overline{\pi}_{\mathrm{B}}(\theta\mid S_n)$ place their posterior mass. 

While ABC methods and standard BSL methods can deliver unreliable inferences when the model is misspecified, BSL can be readily altered to accommodate model misspecification. So far, two approaches have been suggested for accommodating misspecification in the BSL framework. The first suggestion was given in \cite{frazier2019robust} and proposes a robust BSL (R-BSL) method that delivers reliable posterior inferences even in highly misspecified models. When the model is misspecified, there do not exist any values of $\theta\in\Theta$ such that $S_n(z)$ can be made close to the observed summaries $S_n(y)$. R-BSL rectifies this issue by introducing auxiliary parameters into the synthetic likelihood in such a way that the observed summaries $S_n(y)$ can always be matched by the simulated summaries. This then requires conducting posterior inference over a larger set of model unknowns, which includes both the original model parameter and the additional auxiliary parameters, but simultaneously allows us to `learn about' which summaries are misspecified as well. 

An alternative framework for dealing with model misspecification in BSL is the approach of \cite{frazier2021synthetic}, which relies on a two-step approach to accommodate model misspecification. First, a naive BSL posterior is constructed that places posterior mass on values of $\theta$ where the simulated summaries are closest to matching the observed summarise $S_n(y)$. In the second step, this naive posterior is modified to ensure it has certain optimal theoretical properties. 

Consequently, even if the model is misspecified, robust BSL methods can deliver reliable Bayesian inferences in intractable models. In contrast, if the model is misspecified, the ABC posterior is known to display non-standard posterior behavior and its uncertainty quantification is unreliable (\citealp{frazier2020model}).

\section*{Acknowledgements}
David Frazier was supported by the Australian Research Council's Discovery Early Career Researcher Award funding scheme (DE200101070). Christopher Drovandi was supported by an Australian Research Council Future Fellowship (FT210100260).  

\spacingset{1.0} 
{\footnotesize
\bibliographystyle{chicago}
\bibliography{refs_mispec_bsl}
}



\end{document}